\documentclass[reqno]{article}
\usepackage{amsmath,amssymb}

\newcommand{\ba}{\begin{eqnarray}}
\newcommand{\ea}{\end{eqnarray}}
\newcommand{\nl }{ \nonumber  }

\newcommand{\ul}{\underline}
\newcommand{\p}{\partial}

\newcommand{\h}{\hspace{.5cm}}

\title{Probe Branes Dynamics in Nonconstant Background Fields}

\bigskip
\author{Plamen Bozhilov \\
\small\it Department of Theoretical and Applied Physics,\\[-1.mm]
\small\it Shoumen University, 9700 Shoumen, Bulgaria  \\[-1.mm] \small\it
email: p.bozhilov@shu-bg.net}
\date{ }

\begin{document}

\maketitle
\bigskip

\begin{abstract}
We consider probe p-branes dynamics in string theory backgrounds of
general type. We use an action, which interpolates between Nambu-Goto and
Polyakov type actions. This allows us to give a unified description for
the tensile and tensionless branes. Firstly, we perform our analysis in
the frequently used static gauge. Then, we obtain exact brane solutions in
more general gauges. The same approach is used to study the Dirichlet
p-brane dynamics and here exact solutions are also found. As an
illustration, we apply our results to the brane world scenario in the
framework of the mirage cosmology approach.

\end{abstract}

\section{Introduction}
The classical $p$-brane is a $p$-dimensional relativistic object, which
evolving in the space-time, describes a $(p+1)$-dimensional worldvolume.
In this terminology, $p=0$ corresponds to a point particle, $p=1$
corresponds to a string, $p=2$ corresponds to a membrane and so on.

Every $p$-brane characterizes by its tension $T_p$ with dimension of
$(mass)^{p+1}$. When the tension $T_p = 0$, the $p$-brane is called
tensionless one. This relationship between the tensionless branes and the
tensile ones generalizes the correspondence between massless and massive
particles for the case of extended objects. Thus, the tensionless branes
may be viewed as a high-energy limit of the tensile ones.

The point particle ($0$-brane) couples to 1-form gauge field (the
electro-magnetic potential). Generalizing to arbitrary $p$, one obtains
that the $p$-branes couple to $(p+1)$-form gauge potentials.

As is known, there exist five consistent superstring theories in ten
dimensions: Type IIA with $N=2$ non-chiral supersymmetry, type IIB with
$N=2$ chiral supersymmetry, type I with $N=1$ supersymmetry and gauge
symmetry $SO(32)$ and heterotic strings with $N=1$ supersymmetry and
$SO(32)$ or $E_8\times E_8$ gauge symmetry.

The superstring dynamics unify all fundamental interactions between the
elementary particles, including gravity, at super high energies. The
$p$-branes arise naturally in the superstring theory, because there exist
exact brane solutions of the superstring effective equations of motion.
The $2$-branes and the $5$-branes are the fundamental dynamical objects in
eleven dimensional $M$-theory, which is the strong coupling limit of the
five superstring theories in ten dimensions, and which low energy field
theory limit is the eleven dimensional supergravity. Particular type of
$3$-branes arise in the Randall-Sundrum brane world scenario.

In this talk, we consider the probe p-branes dynamics in string theory
backgrounds of generic type. In the probe brane limit, one neglects the
back-reaction of the brane on the background fields. We use an action,
which interpolates between Nambu-Goto and Polyakov type actions. Thus, we
give a unified description for the tensile and tensionless branes. Our aim
is to obtain the conditions on the background metric and the (p+1)-form
gauge field, under which the {\it nonlinear} brane equations of motion and
constraints can be solved exactly. Firstly, we perform our analysis in the
frequently used static gauge. Then, we obtain exact brane solutions in a
more general gauge. The same approach is used to study the Dirichlet
p-brane dynamics and here exact solutions are also found. As an
illustration, we apply our results to the brane world scenario in the
framework of the mirage cosmology approach.

\section{Probe $p$-Branes Dynamics}
The Nambu-Goto type action for bosonic $p$-brane in a $D$-dimensional
curved space-time with metric tensor $g_{MN}(x)$, interacting with a
background (p+1)-form gauge field $B_{p+1}$ via Wess-Zumino term, can be
written as \ba\nl S_p^{NG}=&-&T_p \int d^{p+1}\xi\Bigl\{ \sqrt{-\det[\p_m
X^M\p_n X^N g_{MN}(X)]}\\ \nl &-&\frac{ \varepsilon^{m_1\ldots
m_{p+1}}}{(p+1)!} \p_{m_1}X^{M_1}\ldots\p_{m_{p+1}}X^{M_{p+1}}
B_{M_1\ldots M_{p+1}}(X)\Bigr\}, \\ \nl &&  x^M=X^M(\xi^m),\h
\p_m=\p/\p\xi^m,\\ \nl && m,n = 0,1,\ldots,p;\h M,N = 0,1,\ldots,D-1.\ea
The corresponding Polyakov type action is given by \ba\nl
S_p^{P}=&-&\frac{T_p}{2}\int d^{p+1}\xi\Bigl\{
\sqrt{-\gamma}\left[\gamma^{mn} \p_m X^M\p_n X^N g_{MN}(X)-(p-1)\right]
\\ \nl
&-&2\frac{ \varepsilon^{m_1\ldots m_{p+1}}}{(p+1)!}
\p_{m_1}X^{M_1}\ldots\p_{m_{p+1}}X^{M_{p+1}} B_{M_1\ldots
M_{p+1}}(X)\Bigr\},\ea where $\gamma$ is the determinant of the auxiliary
worldvolume metric $\gamma_{mn}$, and  $\gamma^{mn}$ is its inverse.

None of these actions is appropriate for description of the tensionless
branes. To have the possibility for unified description of the tensile and
tensionless branes, we will use the action \ba\nl S_p &=&\int
d^{p+1}\xi\Bigl\{\frac{1}{4\lambda^0}\Bigl[g_{MN}\left(X\right)
\left(\p_0-\lambda^i\p_i\right) X^M\left(\p_0-\lambda^j\p_j\right)X^N \\
\nl &-&
\left(2\lambda^0T_p\right)^2\det\left[g_{MN}\left(X\right)\p_iX^M\p_jX^N
\right]\Bigr]\\ \nl &+&T_p B_{M_0\ldots
M_p}(X)\p_0X^{M_0}\ldots\p_pX^{M_p} \Bigr\},\h (i,j = 1,\ldots,p), \ea in
which the limit $T_p\to 0$ may be taken. It can be shown that this action
is classically equivalent to the above two actions. The Lagrange
multipliers $\lambda^0$ and $\lambda^i$ are connected to the lapse
function $N$ and the shift vector $N^i$ as follows: \ba\nl
\left(2\lambda^0T_p\right)^2=N^2,\h \lambda^i = N^i.\ea Varying the action
$S_p$ with respect to $\lambda^m$, one obtains the constraints  \ba\nl
&&g_{MN} \left(\p_0-\lambda^i\p_i\right)X^M
\left(\p_0-\lambda^j\p_j\right)X^N\\ \nl &&+\left(2\lambda^0T_p\right)^2
\det\left[g_{MN}\p_iX^M\p_jX^N \right]=0,\\ \nl
&&g_{MN}\left(\p_0-\lambda^i\p_i\right)X^M \p_jX^N=0.\ea We will work in
the gauge $\lambda^m = constants$, in which the equations of motion take
the form: \ba\nl &&g_{LN}\Bigl[\left(\p_0-\lambda^i\p_i\right)
\left(\p_0-\lambda^j\p_j\right)X^N - \left(2\lambda^0T_p\right)^2
\p_i\left(\ul{G}G^{ij}\p_j X^N\right)\Bigr]\\ \nl
&&+\Gamma_{L,MN}\Bigl[\left(\p_0-\lambda^i\p_i\right)X^M
\left(\p_0-\lambda^j\p_j\right)X^N\\ \nl &&- \left(2\lambda^0T_p\right)^2
\ul{G}G^{ij}\p_i X^M\p_j X^N\Bigr] \\ \nl &&=2\lambda^0T_p H_{LM_0\ldots
M_p}\p_0X^{M_0}\ldots\p_pX^{M_p},\ea where \ba\nl
&&\ul{G}=\det\left(G_{ij}\right)=\det\left(g_{MN}\p_iX^M\p_jX^N \right),\h
H_{p+2}=dB_{p+1},\\ \nl &&\Gamma_{L,MN}=\frac{1}{2}\left(\p_M g_{NL}+\p_N
g_{ML}-\p_L g_{MN}\right).\ea

\subsection{Static Gauge}
In the commonly used {\it static gauge}, we have the following
identification: $X^m(\xi)=\xi^m$. The other part of the coordinates $X^a$
are supposed to be functions only of $\xi^0\equiv\tau$. The typical string
theory backgrounds do not depend on $x^m$, so that \ba\nl \p g_{MN}/\p x^m
=0,\h \p B_{M_0\ldots M_p}/\p x^m =0.\ea Under these conditions, the
action $S_p$ reduces to \ba\nl &&S_p^{SG}=\int d\tau L^{SG}(\tau),\h V_p =
\int d^p\xi,\\ \nl &&L^{SG}=
\frac{V_p}{4\lambda^0}\Bigl[g_{ab}\dot{X^a}\dot{X^b} + 2\left(g_{0a}
-\lambda^i g_{ia}+ 2\lambda^0T_p B_{a1\ldots p} \right)\dot{X^a}\\ \nl &&+
g_{00}-2\lambda^i g_{0i} + \lambda^i\lambda^j g_{ij}- \left(2\lambda^0
T_p\right)^2 \det(g_{ij})+ 4\lambda^0 T_p B_{01\ldots p}\Bigr].\ea To have
finite action, we require the fraction $V_p/\lambda^0$ to be finite. In
the string case ($p=1$) and in conformal gauge, for example, this
corresponds to have the fraction $V_1/\alpha'= 2\pi V_1 T_1$ finite.

 Now, the constraints are: \ba\nl &&g_{ab}\dot{X^a}\dot{X^b} +
2\left(g_{0a}-\lambda^i g_{ia}\right)\dot{X^a}\\ \nl &&+ g_{00}-2\lambda^i
g_{0i} + \lambda^i\lambda^j g_{ij}+\left(2\lambda^0T_p\right)^2
\det(g_{ij})=0,\\ \nl &&g_{ia}\dot{X^a}+g_{i0}-g_{ij}\lambda^j=0.\ea

The Lagrangian $L^{SG}$ does not depend on $\tau$ explicitly, so the
energy $E$ is conserved: \ba\nl &&g_{ab}\dot{X^a}\dot{X^b}-
g_{00}+2\lambda^i g_{0i} - \lambda^i\lambda^j g_{ij}+ \left(2\lambda^0
T_p\right)^2 \det(g_{ij})\\ \nl &&- 4\lambda^0 T_p B_{01\ldots p}=
\frac{4\lambda^0 E}{V_p}=constant.\ea With the help of the constraints, we
can replace this equality by the following one \ba\nl g_{0a}\dot{X^a} +
g_{00}-\lambda^ig_{i0}+2\lambda^0 T_p B_{01\ldots p}=- \frac{2\lambda^0
E}{V_p}.\ea

It can be shown that the equations of motion and all these constraints can
be reduced to the equalities \ba\nl &&g_{ab}\ddot{X^b}
+\Gamma_{a,bc}\dot{X^b}\dot{X^c}+\frac{1}{2}\p_a V^{SG} =
2\p_{[a}\mathcal{A}_{b]}^{SG}\dot{X^b},\\ \nl &&g_{ab}\dot{X^a}\dot{X^b}
+V^{SG}=0,\ea where \ba\nl V^{SG}&=&\left(2\lambda^0 T_p\right)^2
\det(g_{ij})- g_{00}+2\lambda^i g_{0i} - \lambda^i\lambda^j g_{ij}\\ \nl
&-& 4\lambda^0\left(T_p B_{01\ldots p}+E/V_p\right)\\ \nl
\mathcal{A}_{a}^{SG}&=&g_{a0}-\lambda^i g_{ai}+2\lambda^0 T_p B_{a1\ldots
p}.\ea

It turns out that for background fields depending on only one coordinate
$x^a$, we can always integrate these equations, and the solution is
\ba\nl\tau\left(X^a\right)=\tau_0 \pm \int_{X_0^a}^{X^a}
\left(-\frac{V^{SG}}{g_{aa}}\right)^{-1/2}d u.\ea Otherwise, supposing the
metric $g_{ab}$ is a diagonal one, we can rewrite the equations of motion
in the form \ba\nl &&\frac{d}{d\tau}\left(g_{aa}\dot{X}^a\right)^2 +
\dot{X}^a\p_a\left(g_{aa} V^{SG}\right)\\ \nl &&+ \dot{X}^a\sum_{b\ne a}
\left[\p_a\left(\frac{g_{aa}}{g_{bb}}\right)\left(g_{bb}\dot{X}^b\right)^2
- 4\p_{[a}\mathcal{A}_{b]}^{SG}g_{aa}\dot{X}^b\right] = 0.\ea

To find solutions of the above equations without choosing particular
background, we fix all coordinates $X^a$ except one. Then the $exact$
probe brane solution of the equations of motion and constraints is given
again by the same expression for $\tau\left(X^a\right)$.

To find solutions depending on more than one coordinate, we have to impose
further conditions on the background fields. An example of such {\it
sufficient} conditions, which allow us to reduce the order of the
equations of motion by one, is given below (we split the index $a$ in such
a way that $X^r$ is one of the coordinates $X^a$, and $X^{\alpha}$ are the
others): \ba\nl
&&\p_{\alpha}\left(\frac{g_{\alpha\alpha}}{g_{aa}}\right)=0,\h
\p_{\alpha}\left(g_{rr}\dot{X}^r\right)^2 = 0,\\ \nl &&
\p_{r}\left(g_{\alpha\alpha}\dot{X}^{\alpha}\right)^2 = 0,\h
\mathcal{A}_{\alpha}^{SG}=\p_{\alpha}f.\ea The result of integrations is
the following \ba\nl \left(g_{\alpha\alpha}\dot{X}^{\alpha}\right)^2 &=&
D_{\alpha} \left(X^{a\ne\alpha}\right) + g_{\alpha\alpha}\left[2\left(
\mathcal{A}_{r}^{SG}-\p_r f\right)\dot{X}^r - V^{SG}\right]\\ \nl &=&
E_{\alpha} \left(X^{\beta}\right),\\ \nl \left(g_{rr}\dot{Z}^{r}\right)^2
&=& g_{rr}\left\{\left(n_{\alpha}-1\right) V^{SG} -
\sum_{\alpha}\frac{D_{\alpha}}{g_{\alpha\alpha}}\right\} +
\left[n_{\alpha}\left(\mathcal{A}_{r}^{SG}-\p_r f\right)\right]^2\\ \nl
&=& E_r\left(X^r\right),\\ \nl \dot{Z}^r &=& \dot{X}^r +
\frac{n_{\alpha}}{g_{rr}} \left(\mathcal{A}_{r}^{SG}-\p_r f\right),\ea
where $D_{\alpha}$, $E_{\alpha}$ and $E_r$ are arbitrary functions of
their arguments, and $n_{\alpha}$ is the number of the coordinates
$X^\alpha$.

Further progress is possible, when working with particular background
configurations.

\subsection{Rotated Gauge}
Now we will repeat our analysis of the probe $p$-brane dynamics in a more
general gauge than the static one. Namely, our ansatz for the coordinates
$X^m(\xi)$ is the following \ba\nl
X^m(\xi)=\Lambda^m_n\xi^n=\Lambda^m_0\tau+\Lambda^m_i\xi^i,
\h\Lambda^m_n=constants.\ea We call this ansatz rotated gauge, because
$\Lambda^m_n\xi^n$ look like rotations in the space described by the
coordinates $\xi^n$. However, there are no restrictions on the parameters
$\Lambda^m_n$. They are {\it arbitrary} constants. For
$\Lambda^m_n=\delta^m_n$, we come back to static gauge.

In the same way as before, one obtains the Lagrangian \ba\nl &&L^{R}=
\frac{V_p}{4\lambda^0}\Bigl\{g_{ab}\dot{X^a}\dot{X^b} +
2\Bigl[\left(\Lambda_0^n-\lambda^i\Lambda_i^n\right)g_{na}+ 2\lambda^0T_p
B_{a}\Bigr]\dot{X^a}\\ \nl &&+
\left(\Lambda_0^n-\lambda^i\Lambda_i^n\right)
\left(\Lambda_0^m-\lambda^j\Lambda_j^m\right)g_{nm}- \left(2\lambda^0
T_p\right)^2 \det(\Lambda_i^n\Lambda_j^m g_{nm})\\ \nl &&+ 4\lambda^0
T_p\Lambda_0^{m}B_{m}\Bigr\},\h\h B_M\equiv B_{Mm_1\ldots
m_p}\Lambda_1^{m_1}\ldots\Lambda_p^{m_p},\ea the constraints \ba\nl
&&g_{ab}\dot{X^a}\dot{X^b} +
2\left(\Lambda_0^n-\lambda^i\Lambda_i^n\right)g_{na}\dot{X^a}+
\left(\Lambda_0^n-\lambda^i\Lambda_i^n\right) \\ \nl &&\times
\left(\Lambda_0^m-\lambda^j\Lambda_j^m\right)g_{nm}+\left(2\lambda^0
T_p\right)^2 \det(\Lambda_i^n\Lambda_j^m g_{nm})=0,\\ \nl
&&\Lambda_i^n\left[g_{na}\dot{X^a}+
\left(\Lambda_0^m-\lambda^j\Lambda_j^m\right)g_{nm}\right]=0,\ea the
conserved energy \ba\nl \Lambda_0^n\left[g_{na}\dot{X^a}+
\left(\Lambda_0^m-\lambda^j\Lambda_j^m\right)g_{nm}
+2\lambda^0T_pB_n\right]=-\frac{2\lambda^0 E}{V_p},\ea and finally, the
equations of motion and the effective constraint \ba\nl &&g_{ab}\ddot{X^b}
+\Gamma_{a,bc}\dot{X^b}\dot{X^c}+\frac{1}{2}\p_a V^{R} =
2\p_{[a}\mathcal{A}_{b]}^{R}\dot{X^b},\\ \nl &&g_{ab}\dot{X^a}\dot{X^b}
+V^{R}=0.\ea We see that these equalities have the same form as in static
gauge, but with $V^{SG}$, $\mathcal{A}_{a}^{SG}$ replaced with $V^{R}$,
$\mathcal{A}_{a}^{R}$, where \ba\nl V^{R}&=&\left(2\lambda^0 T_p\right)^2
\det(\Lambda_i^n\Lambda_j^m g_{nm})
-\left(\Lambda_0^n-\lambda^i\Lambda_i^n\right)\\ \nl &\times&
\left(\Lambda_0^m-\lambda^j\Lambda_j^m\right)g_{nm}-
4\lambda^0\left(T_p\Lambda_0^{m}B_{m}+E/V_p\right),\\ \nl
\mathcal{A}_{a}^{R}&=&\left(\Lambda_0^n-\lambda^i\Lambda_i^n\right)g_{an}
+2\lambda^0 T_p B_{a}.\ea Therefore, the corresponding {\it exact}
solution, depending on one of the coordinates $X^a$ will be
\ba\nl\tau\left(X^a\right)=\tau_0 \pm \int_{X_0^a}^{X^a}
\left(-\frac{V^{R}}{g_{aa}}\right)^{-1/2}d u.\ea

\section{Probe D$p$-Branes Dynamics}
The Dirichlet $p$-branes are extended objects on which the open strings
can end. On the other hand, they are solutions of the superstring
effective equations of motion, carrying Ramond-Ramond charges.

The Dirac-Born-Infeld type action for the bosonic part of the
super-D$p$-brane in a $D$-dimensional curved space-time with metric tensor
$g_{MN}(x)$, interacting with a background (p+1)-form Ramond-Ramond gauge
field $C_{p+1}$ via Wess-Zumino term, can be written as \ba\nl
S&=&-T_{D}\int d^{p+1}\xi \Bigl\{e^{-a(p,D)\Phi}\sqrt{-\det\left(G_{mn} +
B_{mn} + 2\pi\alpha'F_{mn}\right)}\\ \nl &-&\frac{ \varepsilon^{m_1\ldots
m_{p+1}}}{(p+1)!} \p_{m_1}X^{M_1}\ldots\p_{m_{p+1}}X^{M_{p+1}}
C_{M_1\ldots M_{p+1}}\Bigr\}.\ea $T_D$=$(2\pi)^{-(p-1)/2}g_s^{-1}T_p$ is
the D-brane tension, $g_s$ = $\exp\langle\Phi\rangle$ is the string
coupling expressed by the dilaton vacuum expectation value
$\langle\Phi\rangle$ and $2\pi\alpha'$ is the inverse string tension.
$G_{mn}= \p_m X^M\p_n X^N g_{MN}(X)$, $B_{mn}= \p_m X^M\p_n X^N b_{MN}(X)$
and $\Phi(X)$ are the pullbacks of the background metric, antisymmetric
tensor and dilaton to the D$p$-brane worldvolume, while $F_{mn}(\xi)$ is
the field strength of the worldvolume $U(1)$ gauge field $A_m(\xi)$. The
parameter $a(p,D)$ depends on the brane and space-time dimensions $p$ and
$D$ respectively.

To be able to take the limit $T_D\to 0$, we will work with the classically
equivalent action \ba\nl S_D&=&\int d^{p+1}\xi
\frac{e^{-a\Phi}}{4\lambda^0}\Bigl[G_{00}-2\lambda^i G_{0i} +
\left(\lambda^i\lambda^j-\kappa^i\kappa^j\right)G_{ij}\\ \nl
&-&\left(2\lambda^0T_D\right)^2\det(G_{ij}) +2\kappa^i\left(
\mathcal{F}_{0i}-\lambda^j\mathcal{F}_{ji}\right)\\ \nl
&+&4\lambda^0T_De^{a\Phi} C_{M_0\ldots
M_p}\p_0X^{M_0}\ldots\p_pX^{M_p}\Bigr],\\ \nl &&\mathcal{F}_{mn}=B_{mn} +
2\pi\alpha'F_{mn}.\ea Here additional Lagrange multipliers $\kappa^i$ are
introduced, in order to linearize the quadratic term \ba\nl
\left(\mathcal{F}_{0i}-\lambda^k\mathcal{F}_{ki}\right)
\left(G^{-1}\right)^{ij}
\left(\mathcal{F}_{0j}-\lambda^l\mathcal{F}_{lj}\right)\ea in the action.

For simplicity, we restrict our considerations to constant dilaton
$\Phi=\Phi_0$ and constant electro-magnetic field $F_{mn}$ on the
D$p$-brane worldvolume.

\subsection{Static Gauge}
In static gauge, the reduced Lagrangian is given by \ba\nl L_D^{SG}&=&
\frac{V_p e^{-a\Phi_0}}{4\lambda^0} \Bigl[g_{ab}\dot{X^a}\dot{X^b}+
g_{00}-2\lambda^i g_{0i} +
\left(\lambda^i\lambda^j-\kappa^i\kappa^j\right)g_{ij}\\ \nl &+&
2\left(g_{0a} -\lambda^i g_{ia}+ 2\lambda^0T_D e^{a\Phi_0}C_{a1\ldots p}
+\kappa^i b_{ai}\right)\dot{X^a}\\ \nl &-& \left(2\lambda^0 T_D\right)^2
\det(g_{ij})+ 4\lambda^0 T_D e^{a\Phi_0}C_{01\ldots p}\\ \nl &+&2\kappa^i
\left(b_{0i}-\lambda^j b_{ji}\right) +4\pi\alpha'\kappa^i
\left(F_{0i}-\lambda^j F_{ji}\right)\Bigr].\ea

Using the same approach as before, we obtain the equations of motion and
the effective constraint, which have the previous form \ba\nl
&&g_{ab}\ddot{X^b} +\Gamma_{a,bc}\dot{X^b}\dot{X^c}+\frac{1}{2}\p_a
V_D^{SG} = 2\p_{[a}\mathcal{A}_{b]}^{DS}\dot{X^b},\\ \nl
&&g_{ab}\dot{X^a}\dot{X^b} +V_D^{SG}=0,\ea where now \ba\nl V_D^{SG}&=&
\left(2\lambda^0 T_D\right)^2 \det(g_{ij})- g_{00}+2\lambda^i g_{0i} -
\left(\lambda^i\lambda^j-\kappa^i\kappa^j\right)g_{ij}\\ \nl &-&4\lambda^0
e^{a\Phi_0}\left(T_D C_{01\ldots p}+E/V_p\right)\\ \nl &-&2\kappa^i
\left(b_{0i}-\lambda^j b_{ji}\right) -4\pi\alpha'\kappa^i
\left(F_{0i}-\lambda^j F_{ji}\right),\\ \nl
\mathcal{A}_{a}^{DS}&=&g_{a0}-\lambda^i g_{ai}+2\lambda^0 T_D e^{a\Phi_0}
C_{a1\ldots p}+\kappa^i b_{ai}.\ea

The corresponding exact solution, depending on one of the coordinates
$X^a$, is \ba\nl\tau\left(X^a\right)=\tau_0 \pm \int_{X_0^a}^{X^a}
\left(-\frac{V_D^{SG}}{g_{aa}}\right)^{-1/2}d u.\ea

\subsection{Rotated Gauge}
In the same way, one obtains the exact probe D$p$-brane solution in this
gauge. It is given by the above equality for $\tau\left(X^a\right)$, where
instead of $V_D^{SG}$ we have to use the potential \ba\nl
V_D^{R}&=&\left(2\lambda^0 T_D\right)^2 \det(\Lambda_i^n\Lambda_j^m
g_{nm}) -\Bigl[\left(\Lambda_0^n-\lambda^i\Lambda_i^n\right)
\left(\Lambda_0^m-\lambda^j\Lambda_j^m\right)\\ \nl
&-&\kappa^i\kappa^j\Lambda_i^n\Lambda_j^m\Bigr] g_{nm}- 4\lambda^0
e^{a\Phi_0} \left(T_D\Lambda_0^{m}C_{m}+E/V_p\right)\\ \nl
&-&2\kappa^i\Lambda_i^m\left(\Lambda_0^n-\lambda^j\Lambda_j^n\right)b_{nm}
-4\pi\alpha'\kappa^i \left(F_{0i}-\lambda^j F_{ji}\right),\\ \nl
&&C_m=C_{mm_1\ldots m_p}\Lambda_1^{m_1}\ldots\Lambda_p^{m_p}.\ea

\section{Application to Mirage Cosmology}
The idea of the mirage cosmology approach \cite{K99_1} - \cite{TBS02} to
the brane world model is the following \cite{KK99}. The motion of the
brane universe in a curved higher dimensional bulk space induces a
cosmological evolution on the universe brane that is indistinguishable
from a similar one induced by matter density on the brane. It can be shown
that the motion of the probe brane in ambient space induces cosmological
expansion or contraction on our universe simulating various kinds of
"matter" or a cosmological constant (inflation).

There are two steps in the procedure:
\begin{enumerate}
\item {Determine the probe brane motion by solving the worldvolume field
equations.}
\item {Determine the induced metric on the brane which now becomes an
implicit function of time. This gives a cosmological evolution in the
induced brane metric. This cosmological evolution can be reinterpreted in
terms of cosmological "mirage" energy densities on the brane via a
Friedman-like equation for the scale factor $\mathbf{a}$}:
\end{enumerate}
\ba\nl \left( \frac{1}{\mathbf{a}}\frac{d\mathbf{a}}{d\eta}\right)^2
=\frac{8\pi}{3}\rho_{eff}.\ea In addition, one defines the effective
pressure $p_{eff}$ trough the equality \ba\nl
\frac{1}{\mathbf{a}}\frac{d^2\mathbf{a}}{d\eta^2}
=-\frac{4\pi}{3}\left(\rho_{eff}+3p_{eff}\right).\ea

We showed already how the brane equations of motion and constraints can be
solved exactly. The induced line element on the brane is \ba\nl
ds^2=G_{mn}dx^mdx^n = G_{00}(dx^0)^2 + 2G_{0j}dx^0dx^j +
G_{ij}dx^idx^j.\ea Introducing the cosmic time $\eta$ by the relation
\ba\nl d\eta=\sqrt{-G_{00}}dx^0,\ea one obtains \ba\nl ds^2=-d\eta^2 +
2\frac{G_{0j}}{\sqrt{-G_{00}}}d\eta dx^j + G_{ij}dx^idx^j.\ea

Now, let us give the metrics induced on the brane in the different cases
considered. In static gauge, we have the following $p$-brane metric \ba\nl
&&G_{00}=-\left(2\lambda^0T_p\right)^2\det\left(g_{ij}\right)
+\lambda^i\lambda^jg_{ij},\\ \nl &&G_{0j}=\lambda^ig_{ij},\h
G_{ij}=g_{ij}.\ea The corresponding generalization for the D$p$-brane
metric is given by \ba\nl
G_{00}&=&-\left(2\lambda^0T_D\right)^2\det\left(g_{ij}\right)
+\left(\lambda^i\lambda^j-\kappa^i\kappa^j\right)g_{ij}\\ \nl
&-&2\lambda^i\kappa^j\left(b_{ij}+2\pi\alpha'F_{ij}\right),\\ \nl
G_{0j}&=&\lambda^ig_{ij} +\kappa^i\left(b_{ij}+2\pi\alpha'F_{ij}\right),\h
G_{ij}=g_{ij}.\ea In rotated gauge, the induced $p$-brane metric is \ba\nl
&&G_{00}=-\left(2\lambda^0T_p\right)^2
\det\left(\Lambda_i^m\Lambda_j^ng_{mn}\right)
+\lambda^i\lambda^j\Lambda_i^m\Lambda_j^n g_{mn},\\ \nl
&&G_{0j}=\lambda^i\Lambda_i^m\Lambda_j^ng_{mn},\h
G_{ij}=\Lambda_i^m\Lambda_j^ng_{mn}.\ea In the same gauge, the metric
induced on the D$p$-brane is the following \ba\nl
G_{00}&=&-\left(2\lambda^0T_D\right)^2 \det\left(\Lambda_i^m\Lambda_j^n
g_{mn}\right) +\left(\lambda^i\lambda^j-\kappa^i\kappa^j\right)
\Lambda_i^m\Lambda_j^n g_{mn}\\ \nl &-&2\lambda^i\kappa^j
\left(\Lambda_i^m\Lambda_j^nb_{mn} +2\pi\alpha'F_{ij}\right),\\ \nl
G_{0j}&=&\lambda^i\Lambda_i^m\Lambda_j^ng_{mn}
+\kappa^i\left(\Lambda_i^m\Lambda_j^nb_{mn}+2\pi\alpha'F_{ij}\right),\\
\nl G_{ij}&=&\Lambda_i^m\Lambda_j^ng_{mn}.\ea

We will not consider here any particular mirage cosmology. We just mention
that the obtained results allow us to find exact probe branes solutions in
more general background fields and in more general gauges, than known so
far. Therefore, we can induce more general metrics on the probe brane, and
investigate the corresponding mirage cosmology. For example, we can
consider generalized Kasner type metrics, which appear in the superstring
cosmology \cite{LWC99}.

\medskip

{\bf Acknowledgement.} This work is supported by Shoumen University under
Grant No.005/2002.


\end{document}